\documentclass[%
 reprint,
 amsmath,amssymb,
 aps,
]{revtex4-1}

\usepackage{xcolor}
\usepackage{graphicx}
\usepackage{dcolumn}
\usepackage{bm}

\begin{document}

\preprint{APS/123-QED}


\title{Finite size effects on liquid-solid phase coexistence and the estimation of crystal nucleation barriers}

\author{Antonia Statt$^{1,2}$, Peter Virnau$^1$ and Kurt Binder$^1$}

\affiliation{$^1$Institut f\"ur Physik, Johannes Gutenberg-Universit\"at Mainz,Staudinger Weg 9, 55128 Mainz, Germany\\
$^2$ Graduate School of Excellence Materials Science in Mainz, Staudinger Weg 9, 55128 Mainz, Germany}

\date{\today}
             
\begin{abstract} 
A fluid in equilibrium in a finite volume $V$ with particle number $N$ at 
a density $\rho = N/V$ exceeding the onset density $\rho_f $ of freezing 
may exhibit phase coexistence between a crystalline nucleus and surrounding 
fluid. Using a method suitable for the estimation of the chemical potential 
of dense fluids we obtain the excess free energy due to the surface of the 
crystalline nucleus. There is neither a need to precisely locate the interface 
nor to compute the (anisotropic) interfacial tension. As a test case, a soft 
version of the Asakura-Oosawa model for colloid polymer-mixtures is treated. 
While our analysis is appropriate for crystal nuclei of arbitrary shape,
we find the nucleation barrier to be compatible  with a spherical shape, and 
consistent with classical nucleation theory.
\end{abstract}

\pacs{Valid PACS appear here}

\maketitle

Nucleation of crystals from fluid phases and their subsequent growth is 
one of the most important phase transformations in nature \cite{1,2,3}; 
applications range from ice crystal formation in the atmosphere, to 
metallurgy, nanomaterials, protein crystallization, etc. Despite its 
overwhelming importance, crystal nucleation still is only poorly understood.

For the nucleation of a liquid drop from supersaturated vapor, clearly 
the average nucleus shape is spherical. Only the curvature dependence of 
the interfacial tension \cite{4,5,6,7,8,9} presents a stumbling block for 
the prediction of nucleation barriers. Unlike interfaces between fluid 
phases, the crystal-fluid interface tension $\gamma (\vec{n})$ depends 
on the orientation of the interface normal $\vec{n}$ relative to the 
crystal lattice axes \cite{10,11,12}. For isotropic $\gamma$ the nucleus 
is a sphere of radius $R$ (volume $V= 4 \pi R^3/3)$ and its surface excess 
free energy is $F_{surf} = 4 \pi R^2 \gamma = A_{iso}\gamma V^{2/3}$, with 
$A_{iso} = (36 \pi)^{1/3}$. For crystals the term $A_{iso} \gamma$ is 
replaced by a complicated expression,
\begin{equation}\label{eq1}
F_{surf} (V) = \int \limits _{A_\text{W}} \gamma (\vec{n}) d \vec{s} V^{2/3} \equiv A_\text{W} \bar{\gamma} V^{2/3}\; .
\end{equation}
Here $A_\text{W}$ is the surface area of a unit volume whose shape is derivable 
from $\gamma (\vec{n})$ via the Wulff construction \cite{10,11,12} , and 
the average interface tension $\bar{\gamma}$ is defined as 
$\bar{\gamma} = A^{-1}_\text{W}\int \gamma (\vec{n}) d \vec{s}$.

\begin{figure}
\raggedright
\includegraphics[width=0.47\textwidth]{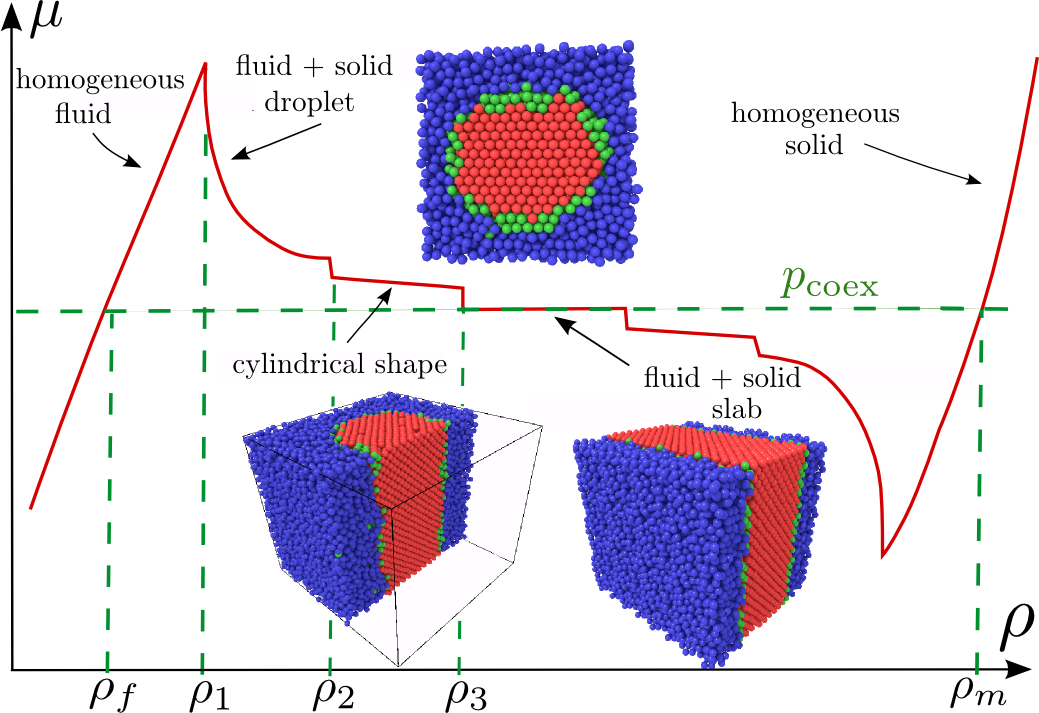}
\caption{\label{fig1} Schematic plot of the chemical potential $\mu$ versus density 
$\rho$ for a system undergoing a liquid-solid transition in a finite box
volume $V_{box}$ with periodic boundary conditions.
(A plot of pressure $p$ vs. $\rho$ would qualitatively 
look just the same). Due to interfacial effects, non-negligible in 
finite systems, the isotherm deviates from $p=p_{coex}$ in the two-phase 
coexistence region, $\rho_f<\rho <\rho_m$. The features in the curve 
(kinks in reality are rounded due to fluctuations) are due to transitions 
between the different states shown in the figure via snapshots of the simulated generalized 
Asakura-Oosawa model. Only the part where the solid phase 
is the minority phase is discussed. For further explanations cf. text.}
\end{figure}

In the classical nucleation theory \cite{1,2,3}, the formation free 
energy of a nucleus is written in terms of volume and surface terms as
\begin{equation}\label{eq2}
\Delta F = - (p_c-p_l) V + F_{surf} (V) \;.
\end{equation}
Here $p_c$ is the pressure in the crystal nucleus and $p_l$ in the (metastable) liquid phase surrounding it. 
In the thermodynamic limit, the configuration with one nucleus on top of the free energy barrier in the 
metastable phase is a saddle point in configuration space. The condition for (unstable) equilibrium,
 $\partial (\Delta F) /\partial V =0$, then yields the critical nucleus volume $V^*$ and barrier $\Delta F^*$,
\begin{equation}\label{eq3}
V^* = \left[\frac {2 A_\text{W} \bar{\gamma}}{3(p_c-p_l)}\right]^3, \quad \Delta F^* = \frac 1 3 A_\text{W} \bar{\gamma} V^{* 2/3} = \frac 1 2 (p_c - p_l)V^*
\end{equation}
Even if $V^*$ is large enough so that correction terms to Eq.~\ref{eq2} can be neglected, 
the application of Eq.~\ref{eq3} is difficult due to lack of knowledge on $A_\text{W}$ and $\bar{\gamma}$. 
This lack of knowledge has hampered the comparison of observed nucleation rates \cite{13,14,15,16} 
(and the barriers extracted from them) and simulations \cite{17,18,19,20} where $\Delta F^*$ was 
estimated directly by biased sampling methods. These comparisons were made for suspensions of 
(hard sphere-like) colloidal particles; the large size of the colloids has the advantage to 
allow direct microscopic observations of crystal-liquid interfaces \cite{21} and nucleation 
events \cite{22,23}. Since kinetic processes for colloids are many orders of magnitude slower 
than for small molecules, colloids are model systems for the study of the liquid-solid 
transition \cite{24,25}, and well suited to separate nucleation from the subsequent crystal growth.

\begin{figure}
\raggedright
\includegraphics[width=0.47\textwidth]{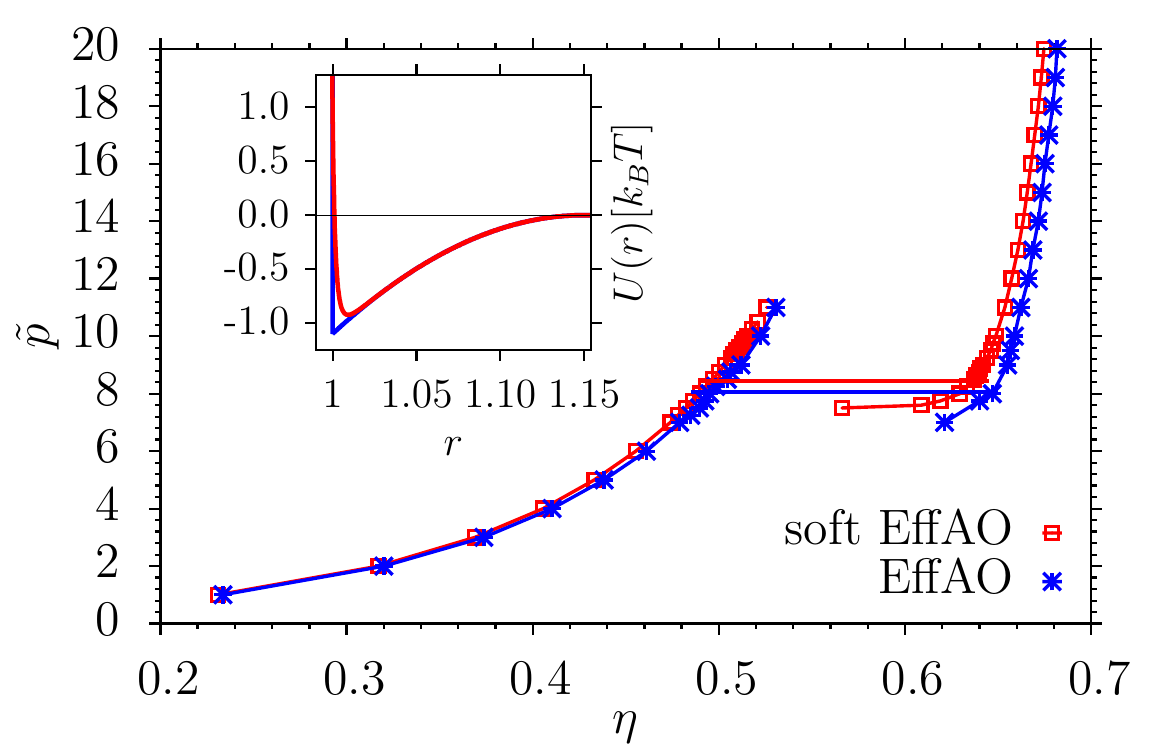}
\caption{\label{fig2} Normalized pressure $\tilde{p} = p \sigma _c^3/k_BT$ 
plotted vs. packing fraction $\eta \equiv \rho \pi \sigma_c^3/6$ of the colloids, 
for the effective AO model (henceforth denoted as Eff AO, asterisks) and its soft version 
(soft Eff AO, squares). Curves are guide to the eye only. These data were obtained from 
simulations of homogeneous liquid and solid (fcc) phases, while the pressure where two-phase 
coexistence occurs was found from the ``interface velocity method'' \cite{37}, 
namely $\tilde{p}=8.44 \pm 0.04$ (soft Eff AO) and $\tilde{p}=8.06 \pm 0.06$ (Eff AO). 
The coexistence packing fractions are $\eta_f = 0.495 $ (1) and $\eta_m = 0.636 $(1) for 
the soft Eff AO case. The insert compares the potentials of the Eff AO 
(which is singular at $r=\sigma_c=1$) and soft Eff AO models.}
\end{figure}

However, to elucidate the persisting discrepancies between simulations 
and experiments one needs to know more about the theoretical nucleation 
barriers: How large must $V^*$ be so that Eq.~\ref{eq3} is a good 
approximation? What is the physical origin of corrections to 
$\Delta F^*$ \{Eq.~\ref{eq3}\} and their magnitude? Is it legitimate to assume
a spherical shape of the nucleus, in spite of its crystalline structure? And so on. 
Understanding the general conditions under which the classical 
description \{Eqs.~\ref{eq2}, \ref{eq3}\} holds will be useful to 
understand liquid-solid transitions in condensed matter in general.

In the present letter, we address these issues, and show how both
 $V^*$ and $\Delta F^*$ can be obtained, considering the equilibrium of 
 the system at fixed finite particle number $N$ in a finite simulation 
 box $V_{box}$. For a suitable range of density $\rho =N/V_{box}$, the 
 equilibrium between the crystalline nucleus and surrounding fluid is 
 perfectly stable.  We explain how both $V^*$ and 
 $p_c-p_l$ can be estimated directly and accurately. Using then 
 $\Delta F^*=(p_c-p_l)V^*/2$ \{Eq.~\ref{eq3}\}, the need of dealing with 
 $\gamma(\bar{n})$ and use of Eq.~\ref{eq1} is bypassed. So we do not need
 to assume anything on the shape of the nucleus.
 
 Thus, the central 
 idea of the present work is to explore the deviations from phase coexistence 
 in the thermodynamic limit (where the chemical potential $\mu = \mu_{coex}$ 
 and the pressure $p = p_{coex} $ for all densities from the onset density 
 of freezing $\rho_f$ to the onset density of melting $\rho_m$) caused by 
 finite size. Thus, the part of the isotherm in Fig.\ref{fig1} corresponding to the 
 homogeneous fluid for finite volume $V_{box}$ exceeds $\rho_f$ and 
 continues up to the ``droplet evaporation condensation transition'' \cite{26} 
 at $\rho_1$, where for the first time a crystalline droplet in the system 
 becomes stable. Note that this transition is a sharp phenomenon only 
 when $V_{box} \rightarrow \infty$ (and then $\rho_1 \rightarrow \rho_f$,
  consistent with the lever rule \cite{27}). At a second special density 
  $\rho_2$ the ``droplet'' changes its shape from compact to cylindrical 
  (stabilized by the periodic boundary conditions).
  At about $\rho = \rho_3$ 
  a slab configuration, separated from the fluid  by two planar interfaces, appears (Fig.1).
  In this region $\mu = \mu_{coex}$ and 
  $p=p_{coex}$ holds true also in the finite system, if
  the linear dimensions in  the directions parallel to the planar interfaces are chosen such that the 
  crystal (at density $\rho_m$) is commensurate without any distortion.
   The analogous behavior for 
  vapor to liquid transitions is well studied \cite{9,28,29,30}. Here 
  we show that the descending part of the $p(\rho)$ and $\mu(\rho)$ 
  isotherms can be used to extract information on $F_{surf},V^*$ and $\Delta F^*$ 
  for the liquid-solid transition as well.

In the snapshots the particles in the fluid region are shown in blue, 
in the crystal in red color, using the averaged Steinhardt local bond order parameters \cite{31,32} 
to distinguish the character of the phases (see Ref.\cite{32} for definitions 
and implementation details). Particles in the interfacial region, for which this 
classification yielded ambiguous results, are shown in green color.
 The face-centered cubic (fcc) packing of the crystal is clearly seen, 
 and the cross section through the ``droplet'' also suggests that the shape may non spherical.

The model of our simulations qualitatively describes colloid polymer 
mixtures \cite{33,34,35,36}. In the Asakura-Oosawa (AO) model \cite{33}, colloids 
are described by hard spheres of diameter $\sigma_c$, polymers as soft spheres 
(which may overlap each other without energy cost) of diameter $\sigma_p$.
Of course, the mutual overlap of colloids and polymers is also strictly forbidden. 
Polymers create the (entropic) depletion attraction between colloids \cite{33}; varying 
the size ratio $q=\sigma_p/\sigma_c$ and the polymer density one can tune the phase 
diagram \cite{34,35,36} and interfacial properties \cite{37,38}. A useful feature of 
this model occurs for $q < q^*=0.154$ \cite{35,39}: then one can integrate 
out the polymer degrees of freedom exactly, and one is left with an effective pairwise 
potential, which is attractive in the range $\sigma_c < r <\sigma_c + \sigma_p$ 
(and zero for $r> \sigma _c + \sigma_p$), but infinitely repulsive for $r < \sigma_c$. 
The strength of the potential of this ``effective'' AO model is controlled by the 
fugacity $z_p$ of the polymers \cite{39} (Fig.~2, insert).

However, it is computationally more convenient to replace the Eff AO model by a 
similar but continuous potential, the soft Eff AO model \cite{39} (Fig.~2, insert). 
For this model the pressure (in the fluid phase) is straightforwardly obtained 
in the simulation from the Virial expression \cite{39,40}, while for the Eff AO model 
due to the discontinuity at $r = \sigma_c$ this is very cumbersome \cite{38}. Fig.~2 shows 
that the variation of $p$ with $\eta$ is very similar for both potentials. 
Since real colloids never are described by hard spheres precisely \cite{41}, 
nor are polymers precisely modeled by ideal soft spheres \cite{42}, a quantitatively 
accurate modeling of real systems anyway cannot be attempted. The soft Eff AO model 
is proposed here as a coarse-grained qualitative model of colloid-polymer mixtures 
which is practically useful in a simulation context.

Using the Virial expression the pressure $p_l$ of the liquid in the region surrounding 
the crystal nucleus in Fig.~1 (far away from the interfacial region) can 
be readily measured, but obtaining $p_c$ inside the nucleus for small nuclei 
is not reliably possible. It is necessary to base the analysis of the two-phase 
equilibrium in $V_{box}$ on the chemical potential $\mu$, because $\mu$ is strictly 
constant in equilibrium also in a spatially inhomogeneous situation. But the 
standard particle insertion method \cite{40,43} does not work at high packing 
fractions $\eta _c$ near $\eta_m$. Thus, we have extended an approach \cite{44} 
to sample the chemical potential of a dense fluid by studying 
a system where walls are present; using a soft wall that reduces the density 
suitably such that there particle insertion works (Fig.~3). Of course, 
it is important to choose $L_z$ large enough so that outside of the range 
of $z$, for which the walls affect the density profile, actually a constant 
density is reached. Fig.~3 demonstrates that in this way the chemical potential 
can be obtained accurately even for $\eta > \eta_f$. The pressure $p$ 
(computed in the region where $\eta(z) = \eta_{bulk} =$ constant) agrees with 
the corresponding bulk data of Fig.~2.

Now we exploit the fact that $\mu$ is constant throughout the system 
also when a crystalline nucleus is present (Fig.~1): the chemical
 potential in the fluid $\mu_f(p_l)$ equals that of the crystal 
nucleus $\mu_c(p_c)$. From  $\mu_c(p_{coex})=\mu_l(p_{coex})= \mu_{coex}$ 
we readily find, using the expansions
\begin{align}
\mu_c(p_c) \approx \mu_{coex} + \frac \pi 6 \frac {1}{\eta_m} (p_c-p_{coex}) \label{eq4} \;, \\
\mu_l(p_l)\approx \mu_{coex} + \frac \pi 6 \frac {1} {\eta_f} (p_l-p_{coex}) \label{eq5}\;,
\end{align}
that $(p_c-p_{coex}) \eta_f = (p_l-p_{coex})\eta_m$. Since we have recorded
 both functions $\mu_l(\eta)$ and $p_l(\eta)$, we also know $\mu_l(p_l)$
  and hence can verify that the data indeed fall in the regime where the 
  linear expansion, Eq.~\ref{eq5}, holds. Finding $\mu_c(p_c)$ via thermodynamic 
  integration (using $\mu_c(p_{coex})= \mu _l (p_{coex})$ as starting point),
   we have verified that Eq.~(\ref{eq4}) also introduces only negligible errors.

\begin{figure}
\raggedright
\includegraphics[width=0.47\textwidth]{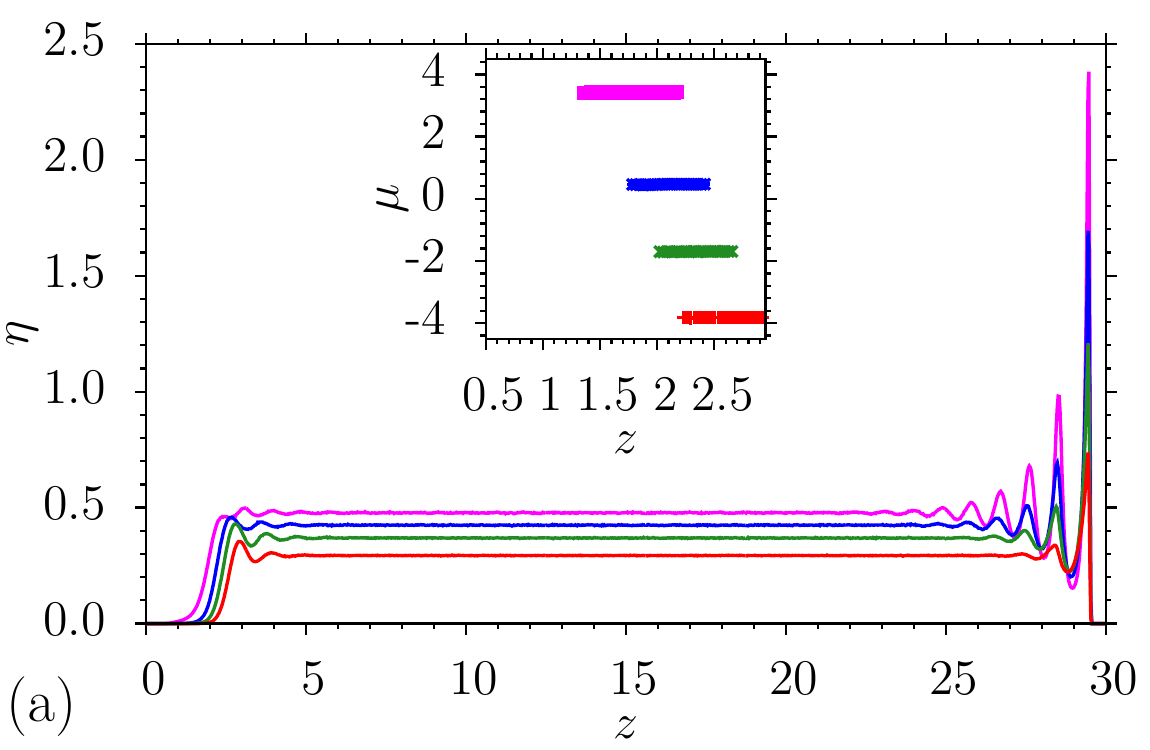}
\includegraphics[width=0.47\textwidth]{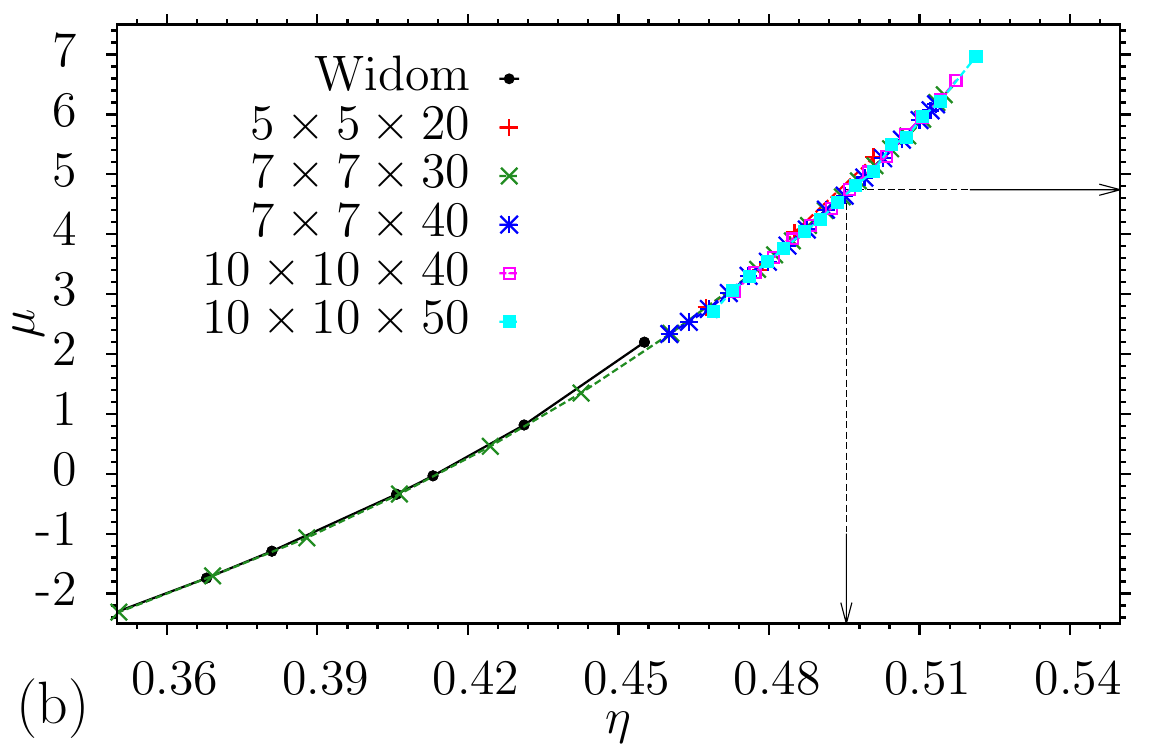}
\caption{\label{fig3} (a) Illustration of the method to compute the chemical potential 
of a very dense fluid, using a $L \times L \times L_z$ slab geometry, with a soft wall 
at $z=0$ and a hard wall at $z=L_z=30$ (lengths being measured in units of $\sigma_c$, $L = 7$, 
and 4 choices of $N$ are used, $N = 750, 950, 1100$ and $1250$, respectively). 
Insert shows $\mu$ in units of $k_BT$ as a function of $z$, for the 4 choices shown, 
over the regions of $z$ where particle insertion works. 
(b) Chemical potential $\mu$ (in units of $k_BT$) plotted vs. $\eta$, for 
different choices of $L$ and $L_z$, as indicated, to show that finite size 
effects are negligible. The data labeled by ``Widom'' at not so large $\eta$ 
are obtained by the standard particle insertion method for homogeneous bulk systems. 
Arrows on abscissa and ordinate indicate $\eta_f$ and $\mu_c(p_{coex})/k_BT$, respectively.}
\end{figure}

The two-phase equilibrium of a crystalline droplet surrounded by fluid has 
been studied for three system sizes, keeping the number of colloids in the 
simulation box fixed (at N = 6000, 8000, and 10 000, respectively) and 
varying $V_{box}$ and hence $\rho=N/V_{box}$. In thermal equilibrium, 
we then have a finite-size variant of the lever rule
\begin{equation}\label{eq6}
\eta V_{box}= \eta_l(p_l)(V_{box}-V^*) + \eta_c(p_c)V^*
\end{equation}
While for $V_{box} \rightarrow \infty$ we would have $p_l=p_c =p_{coex}$ and $\eta_c(p_{coex}) =\eta_m$,
 in the finite system $p_l,p_c$ and the corresponding packing fractions 
 differ from their coexistence values. Initializing the simulation by 
 putting a crystal of about the right volume $V^*$ and about the right 
 choice for $\eta_c(p_c)$ in the box, after a long period of equilibration
 we measure both $p_l$ and $\eta_l(p_l)$ in the fluid region (far away 
 from the crystal) and verify (from the data of the bulk simulation, Fig.~3b) 
 that equilibrium has been reached. Since we know also the chemical potential 
 $(\mu_c (p_c)= \mu_l(p_l)$ is constant), we can obtain $p_c$ and also $\eta_c(p_c)$ 
 and hence Eq.~\ref{eq6} determines $V^*$ unambiguously.

Fig.~4 shows the data for $\Delta p=p_l-p_{coex}$ versus $\eta$. Actually 
when we use the chemical potential $\mu_l(p_l)$ from Fig.~3b and obtain 
$\Delta p$ from Eq.~\ref{eq5}, the data are precisely reproduced, which just 
is a consistency check. 
From simulations determining $\gamma$ for interfaces parallel to 111,110 and 100 planes
~\cite{45} it is found that $\gamma(\vec{n})$ depends only very weakly on $\vec{n}$.
For comparison with classical nucleation theory, we neglect the dependence on $\vec{n}$ and take $\gamma \approx \gamma_{111} \approx \gamma(\vec{n})=1.013$ \cite{45}. 
Assuming a spherical shape  $V^*=4\pi R^{*3}/3$ we find
$\Delta p = (2 \gamma/R^*)/(\eta_m/\eta_f-1)$. Using the observed 
values of $V^*$ one then obtains a prediction for the curves $\Delta p(\eta)$. 
We find that these predicted curves fall slightly below the actual observed data.
They can be brought in good agreement if they are rescaled by a constant factor
of $c=1.07$. This small enhancement can be due to the ratio $A/A_{iso}$ or errors in the
estimation of $\gamma(\vec{n})$.
Unexpectedly, we hence find that for our model of colloid polymer mixtures the assumption of
a spherical nucleus shape works rather well, but it would not be needed to predict the
nucleation barrier. Using Eq.~\ref{eq2}, 
knowledge of $p_c-p_l$ and $V^*$  suffices to predict $\Delta F^*$.
One can expect, however, that significant derivations from spherical nucleus
shape will appear for large $\eta^r_p$ in our model, where the fluid is a vapor-like phase,
and $\gamma(\vec{n})$ will depend more strongly on $\vec{n}$.
Gratifyingly, 
Fig.~4b shows that the three choices for $N$ superimpose to a common curve, so in 
the shown regime finite size effects essentially are negligible.

In summary, we have shown that for the liquid-solid-transition a description 
of nucleation barriers in terms of the classical nucleation theory holds, 
at variance with studies of nucleation with hard sphere-like colloids \cite{13,14,15,16,17,18,19,20,46}.
However, we feel the latter studies are inconclusive, do to their use of too large $\eta_l$
($0.53<\eta_l<0.57$) where the slowing down due to the kinetic prefactor of the nucleation rate
matters \cite{47}. While the range of $\Delta F^*$ in Fig.~4b corresponds to $\eta_l/\eta_f -1 \leq 0.06$
the range of the experiments in Fig.~4b would correspond to $5< \Delta F^* < 10$ only.

Analyzing finite size effects on phase coexistence, 
both $V^*,\; p_l, \; p_c$ and the chemical potential for this stable two-phase 
coexistence in a finite simulation box can be reliably estimated.
The numerical results also clearly show that in the regime where $\Delta F^* \ge 80$
the relation $\Delta F^* \propto {V^*}^{2/3}$ holds precisely, as visible from the fit
in Fig.4b; thus we have verified
that classical theory of homogeneous nucleation for crystals is accurate, in this regime of barriers,
provided one takes into account that the nucleus shape is in general nonspherical.
However, since the two straight lines in Fig.4b almost coincide, the spherical
approximation is shown here to be almost perfect.
Since crystal faces in contact with a dense fluid are frequently atomically rough, 
the spherical approximation is expected to be quite good generally, 
in particular for somewhat  smaller nuclei, for which the nucleation rates also would be larger.

\underline{Acknowledgments}: 
This research was supported by the Deutsche Forschungsgemeinschaft (grant No. VI237/4-3).
We thank the H\"ochstleistungsrechenzentrum Stuttgart (HLRS) and the Zentrum f\"ur 
Datenverarbeitung Mainz for generous grands of computing time at the HERMIT and MOGON
supercomputers.

\begin{figure}
\centering
\includegraphics[width=0.47\textwidth]{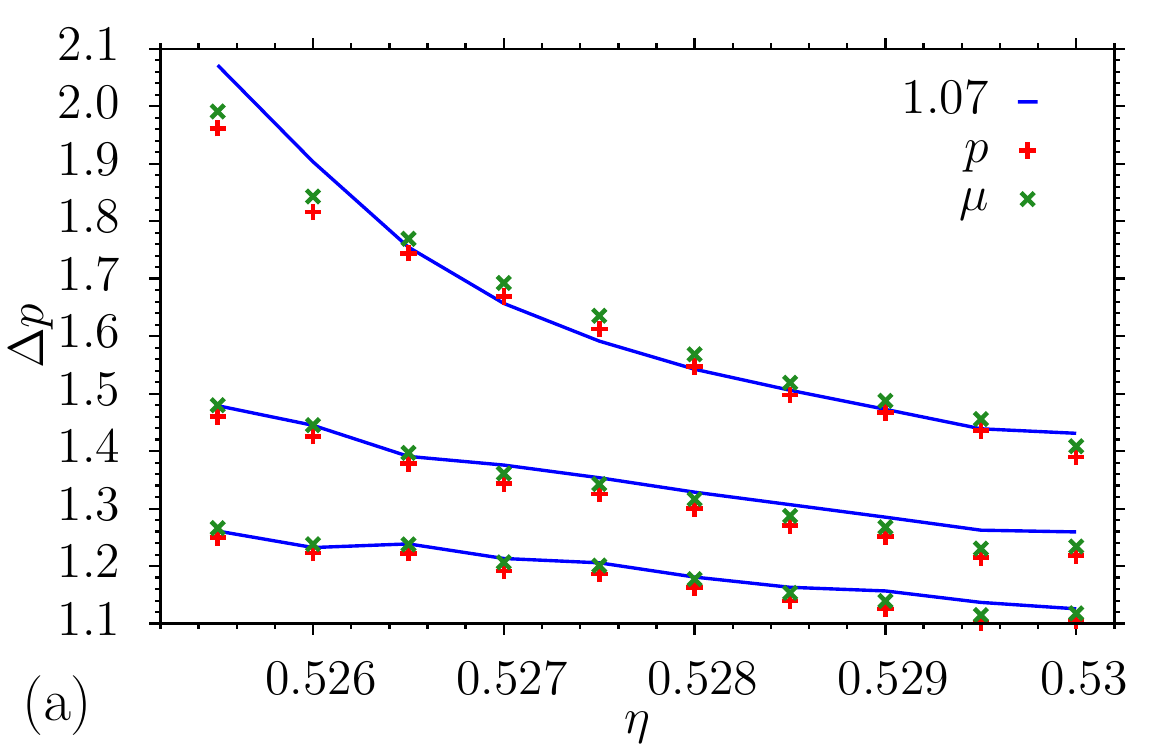}
\includegraphics[width=0.47\textwidth]{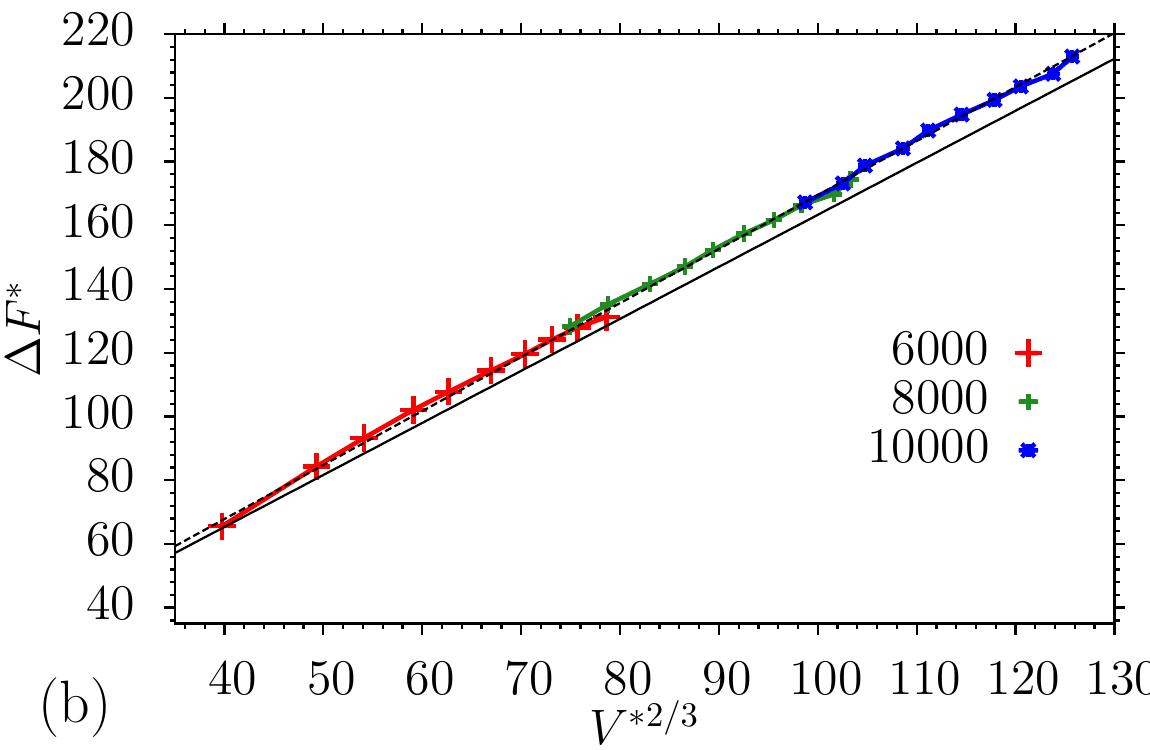}
\caption{\label{fig4} (a) Pressure difference $\Delta p = p_l - p_{coex} $
 between the pressure $p_l$ in a fluid surrounding a crystal nucleus of
 finite size and the coexistence pressure, plotted versus the average packing 
 fraction $\eta$ in the simulation box, for particle number N = 6000, 8000 and 
 10000 (symbols, from bottom to top). Curves show the formula $\Delta p=(2 \gamma/R^*) c/(\eta_m/\eta_f-1)$ with $c=1.07$, 
 extracting $R^*$ from the assumption of a spherical nucleus $(V^*= 4 \pi R^{*3}/3$) and 
 taking $\gamma \approx \tilde{\gamma}_{111} \approx 1.013$ \cite{45}. (b) $\Delta F^*$ computed from 
 $p_l,p_c$ and $V^*$ (using Eqs.~\ref{eq2}, \ref{eq5} and \ref{eq6}) plotted vs. $V^{*2/3}$, straight line
 is Eq.(3) with $A=A_{iso}$ and $\tilde \gamma = \gamma_{111}.$ 
 The broken line is a fit illustrating $\Delta F^* \propto {V^*}^{2/3}$.
 }
\end{figure}

\end{document}